\documentclass[aps,pra,twocolumn,letterpaper,showpacs,superscriptaddress,floatfix,10pt]{revtex4-1}
\usepackage[pdftex]{graphicx}
\graphicspath{ {./images/} }
\usepackage{dcolumn}
\usepackage{bm}
\usepackage{bbm}
\usepackage[usenames, dvipsnames]{color}
\usepackage{comment}
\usepackage[colorlinks, linkcolor=blue]{hyperref}

\usepackage{layouts}

\usepackage[T1]{fontenc}
\usepackage{exscale}
\usepackage{amsbsy}
\usepackage{subfigure}
\usepackage{textcomp}
\usepackage{physics}
\usepackage{mathtools}

\usepackage{amsfonts}
\usepackage{amssymb}
\usepackage{amsmath}
\usepackage{mathtools}
\usepackage{tensor}

\usepackage{amsthm}
\theoremstyle{remark}

\usepackage{multirow}

\usepackage[scr=boondoxo, scrscaled=1.05]{mathalfa}

\usepackage{subfigure}
\usepackage{overpic}

\usepackage{array}
\newcolumntype{L}[1]{>{\raggedright\let\newline\\\arraybackslash\hspace{0pt}}m{#1}}
\newcolumntype{C}[1]{>{\centering\let\newline\\\arraybackslash\hspace{0pt}}m{#1}}
\newcolumntype{R}[1]{>{\raggedleft\let\newline\\\arraybackslash\hspace{0pt}}m{#1}}
\usepackage{tabularx}

\def\KeyWord#1{$\backslash$\IfColor{$\!\!$\textRed{#1}\textBlack}{#1}$\!\!$}

\usepackage{wasysym}

\renewcommand{\d}{\mathrm{d}}
\newcommand{\e}{\mathrm{e}}

\newcommand{\Bt}{{\vec{\theta}}}

\def\Im{\mathrm{Im}}

\def\bra#1{\langle#1|}
\def\ket#1{|#1\rangle}

\begin{document}
\title{Giant energy oscillations mediated by a quasiperiodically driven qubit}

\author{Dominik Vuina}
\email{dominikv@bu.edu}
\affiliation{Department of Physics, Boston University, Boston, Massachusetts 02215, USA}
\author{David M. Long}
\affiliation{Department of Physics, Boston University, Boston, Massachusetts 02215, USA}
\author{Philip J. D. Crowley} 
\affiliation{Department of Physics, Massachusetts Institute of Technology, Cambridge, Massachusetts 02139, USA}
\affiliation{Department of Physics, Harvard University, Cambridge, Massachusetts 02138, USA}
\author{Anushya Chandran}
\affiliation{Department of Physics, Boston University, Boston, Massachusetts 02215, USA}

\begin{abstract}    
    A qubit driven by two incommensurate frequencies can mediate a quantised average energy current in the adiabatic limit. 
    We show that non-adiabatic processes result in reversals of the energy current and corresponding oscillations in the net energy transferred between the drives. 
    The oscillations are bounded but \emph{giant}---much larger than the qubit energy splitting. 
    A Landau-Zener analysis predicts that the timescale of the oscillations is exponentially large in the period of the drives. 
    However, numerical analysis reveals that this timescale is not a monotonic function of the period, and has increasing sub-structure as the adiabatic limit is approached. 
    We show that this non-monotonicity arises from interference effects between subsequent Landau-Zener transitions.
    Giant energy oscillations should be observable in near-term experiments with nitrogen-vacancy centers.
\end{abstract}

\maketitle

\section{Introduction}\label{sec:intro}
Strong driving by multiple incommensurate frequencies can enrich the physical properties of qubits and lattice models.
In particular, each frequency gives rise to an additional \emph{synthetic} lattice dimension~\cite{synthetic_dim, PhysRevX.7.041008_Ivar, PhysRevA.7.2203_synthetic_dim, Verdeny_2016_synthetic_dim, PhysRevB.97.134303_nonad_energy_pump, PhysRevB.98.220509_majorana, PhysRevB.99.064306_PC_IM_AC}. Topological invariants in the synthetic space then manifest as non-equilibrium quantised responses in the driven system. Examples in the adiabatic regime include the well-known Thouless pump~\cite{PhysRevB.27.6083_thouless_original,thouless_pump_bloch_experiment,nakajima_thouless_pump_experiment, PhysRevLett.120.150601_kol_energy_pump}, the qubit energy pump~\cite{PhysRevX.7.041008_Ivar, PhysRevB.99.064306_PC_IM_AC, PhysRevResearch.2.043411_FN_GR_MR, PhysRevLett.126.106805_DL_PC_AC, PhysRevLett.125.100601_PC_IM_AC, PhysRevLett.125.160505_EB_PC_AC_AS,PhysRevB.99.094311_FN_IM_GR, PhysRevLett.128.183602_boosting, PhysRevB.106.L140304_burn_glass}, and non-adiabatic charge~\cite{PhysRevX.6.021013_AFAI, PhysRevB.101.041403_AFAI_charge, PhysRevB.99.195133_AFI, PhysRevX.3.031005_AFI, runder_lindner_AFAI, PhysRevB.96.155118_AFAI, PhysRevB.95.195128_dynamical_symmetries,AFAI_experiment, PhysRevLett.129.256401_AFAI} or energy~\cite{PhysRevLett.126.106805_non_ad_energy_pumps, PhysRevB.106.144203_coupled_layers, PhysRevLett.127.166804_non-adiabtic_energy_pump, PhysRevB.104.224301_nonad_energy_pump} pumps.
Energy pumps can be used to prepare highly non-classical states of light~\cite{PhysRevLett.128.183602_boosting}, which have applications in quantum metrology and error correction~\cite{Caves1981,Giovannetti2004,Cable2007,Demkowicz2015,Brune1996,Chuang1997,Vlastakis2013,Mirrahimi2014,Reagor2016,Xiao2019,Terhal2020,Ma2021}. In many-body settings, quasiperiodic driving can result in emergent dynamical symmetries~\cite{PhysRevX.10.021032_dynamical_phases,  PhysRevLett.123.016806_dynamical_symm, PhysRevB.105.115117_dynamical_symm, dynamical_symm_exp}, time crystals~\cite{PhysRevLett.117.090402_time_crystals, PhysRevLett.118.030401_time_crystals, Zhang_2017_time_crystal, khemani2019brief_time_crystals} and protected edge qubits~\cite{PhysRevB.98.220509_majorana, PhysRevResearch.3.023108_majorana}.

\begin{figure}
\centering
    \includegraphics[width=\linewidth]{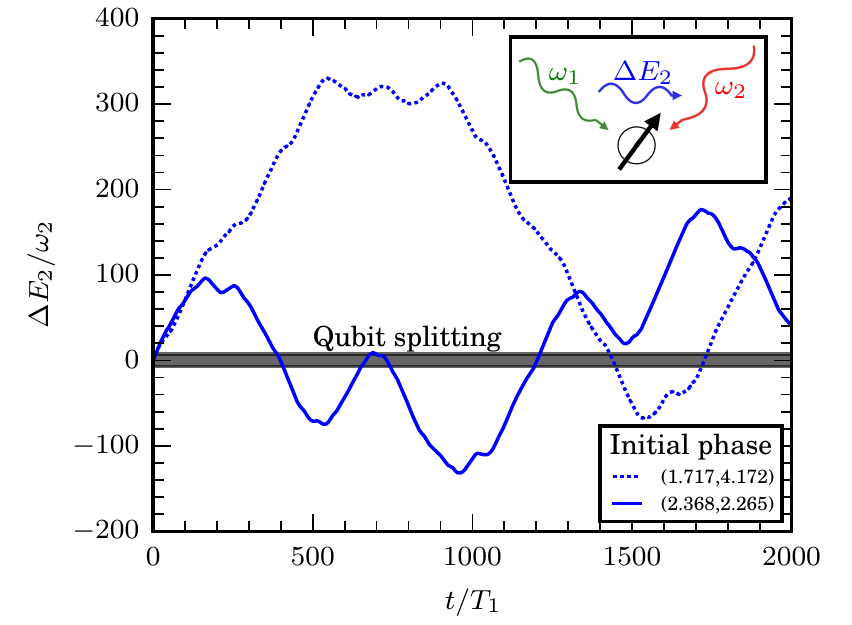}
    \caption{A qubit driven by two incommensurate frequencies \(\omega_1\) and \(\omega_2\) mediates an energy current between the drives in the adiabatic limit (inset). At long times, the energy of each drive, \(\Delta E_{1,2}\) exhibits giant oscillations---much larger than the maximal energy splitting of the qubit (shaded region). \emph{Parameters:} $B_0=2$, $A=1$, $T_1 = 25$ in the BHZ model Eq.~\eqref{eq:normal_bhz_model}. \(\Delta E_2\) is calculated by integrating the spin lock fidelity (which is a proxy for the energy current mediated by the qubit---Sec.~\ref{sec:cur_proxy}).}
    \label{fig:2_tone_driven_qubit_sketch}
\end{figure}

This article concerns the qubit energy pump. A qubit driven by two incommensurate drives has a two-dimensional synthetic space. When the driving is slow and the synthetic lattice exhibits a quantum Hall effect~\cite{PhysRevLett.45.494_klitzing, RevModPhys.58.519_hall_effect_review}, the qubit mediates a time-averaged energy current between the drives when prepared in the instantaneous ground state (\(\hbar=1\))~\cite{PhysRevX.7.041008_Ivar}:
\begin{equation}\label{eq:pump_drive1_to2}
    [P_{1\to 2}]_t = C \frac{\omega_1 \omega_2}{2\pi} + O(t^{-1}).
\end{equation}
Above, $C$ is the Chern number of the synthetic ground state band, $\omega_{1,2}$ are the frequencies of the two drives, and \([\cdot]_t\) denotes an average up to time \(t\). The long time average current $C\omega_1\omega_2/2\pi$ corresponds to a transfer of $C$ energy quanta, \(\omega_2\), of drive two per period of drive one, \(T_1 = 2\pi/\omega_1\). As the Chern number is a topological invariant, the energy current is robust---insensitive to details of the driving protocol.

The energy current in Eq.~\eqref{eq:pump_drive1_to2} relies on the adiabatic evolution of the qubit. The magnitude of non-adiabatic effects are controlled by the \emph{adiabatic parameter} \(\delta\)---the ratio of the squared instantaneous gap to the rate of change of the Hamiltonian. In the two-level models we consider, this will take the form $\delta \approx B /\omega_{1}$, where \(B\) is the amplitude of the drive. The adiabatic theorem~\cite{bachmann2020adiabatic_ad_theorem, landau19322,majorana1932atomi,stuckelberg1932theorie, zener1932non, PhysRevLett.119.060201_ad_theorem} indicates that adiabaticity is violated on timescales \(\tau\) exponentially large in \(\delta\), $\omega_{1}\tau \approx e^{2\pi \delta}$.

The behaviour of the energy current beyond the timescale $\tau$ has not previously been studied. We find that the energy current quasiperiodically oscillates on the timescale $\tau$ indefinitely. In concert, the qubit quasiperiodically oscillates between the instantaneous ground and excited states. As $\tau$ is exponentially large in $\delta$, slowly oscillating energy currents lead to \emph{giant} oscillations in the net energy transferred between the drives---far exceeding the qubit energy splitting $B$ (Fig.~\ref{fig:2_tone_driven_qubit_sketch}). Furthermore, $\tau$ does not increase monotonically with the adiabatic parameter $\delta$. While the coarse-grained behavior is exponential in $\delta$, there are large variations about the theoretical prediction.

The oscillations of the energy current follow from the observation that the qubit spectrum is pure-point~\cite{PhysRevB.99.064306_PC_IM_AC,barata2000formal,blekher1992floquet,jauslin1992generalized,gentile2004pure,jauslin1991spectral}---all observables behave as if the spectrum is discrete, and thus oscillate quasiperiodically. The exponential scaling of $\tau$ is a consequence of the non-adiabatic effects in qubit dynamics. Through the use of the \emph{adiabatic impulse model} (AIM) we are able to numerically simulate the very long timescales needed to observe these oscillations (Sec.~\ref{sec:num_model_TM}).

The fine structure in $\tau(\delta)$ is a numerical observation (Sec.~\ref{sec:period_od_slosh}). We show this stems from the interference effects in instantaneous state population dynamics of the qubit. Indeed, adding sufficiently strong dephasing to the qubit dynamics results in \(\tau\) becoming monotonic in \(\delta\) (Sec.~\ref{sec:decoherence}).

\section{Energy pumping}\label{sec:background}
\subsection{Model} \label{sec:model}
    We consider energy pumping in a \emph{skew-BHZ model}~\footnote{Named skew due to the anisotropy of the drive fields---the \(B_x\) component differs in amplitude to \(B_y\) and \(B_z\). Note that the results presented in this work are independent of the skewness parameter \(A\). This model was introduced as a computational tool amenable to treatment with AIM.} of a qubit driven by two circularly polarized fields. The time dependent Hamiltonian is given by
    \begin{equation}\label{eq:normal_bhz_model}
        H(t) = \frac{1}{2}\vec{B}(\vec{\theta}_t)\cdot \vec{\sigma},
    \end{equation}
    where $\vec{\theta}_t \equiv (\theta_{1t}, \theta_{2t})$ and
    \begin{equation}
         \vec{B}(\vec{\theta}_t) = B_0 \begin{pmatrix} A\sin{\theta_{1t}} \\
         \sin{\theta_{2t}} \\
         1-\cos{\theta_{1t}} - \cos{\theta_{2t}}
         \end{pmatrix}.
    \end{equation}
    The drive phases advance with angular frequencies \(\omega_1\) and \(\omega_2\) respectively, so that \(\Bt_t = \vec{\omega}t +\Bt_0\), where $\vec{\theta}_0$ are initial phases of the drive. We fix $\omega_2/\omega_1 = (1+\sqrt{5})/2$ to be the golden ratio. 
    
    The dimensionless parameter $A$ controls the skewness of the model---the relative amplitude of the external fields generated by the drive. $A=1$ is the well studied BHZ model in Refs.~\cite{PhysRevX.7.041008_Ivar, PhysRevB.99.064306_PC_IM_AC, PhysRevResearch.2.043411_FN_GR_MR, PhysRevLett.126.106805_DL_PC_AC, PhysRevLett.125.100601_PC_IM_AC, PhysRevLett.125.160505_EB_PC_AC_AS,PhysRevB.99.094311_FN_IM_GR, PhysRevLett.128.183602_boosting, PhysRevB.106.L140304_burn_glass, PhysRevB.74.085308_bhz,bernevig2006quantum}. We will consider the $A \gg 1$ limit as this is amenable to treatment with the adiabatic impulse model (Sec.~\ref{sec:num_model_TM}). 
    
    The instantaneous eigenstates of the model are given by the states anti-aligned and aligned with the driving field
    \begin{equation}\label{eq:inst_eigenstates}
        H\ket{\phi_{\pm}(\vec{\theta}_t)} = \pm \frac{1}{2} |\vec{B}(\vec{\theta}_t)|\ket{\phi_{\pm}(\vec{\theta}_t)}.
    \end{equation}
    
\subsection{Adiabatic limit}\label{sec:strict_ad_lim}

    In the adiabatic limit ($\vec{\omega}\to 0$) time evolved states follow instantaneous eigenstates of the model~\eqref{eq:inst_eigenstates} 
    \begin{multline}\label{eq:inst_ground_time_evo}
        \ket{\psi(0)} = c_-(0)\ket{\phi_{-}(\Bt_0)} + c_+(0)\ket{\phi_{+}(\Bt_0)} \\
        \implies
        \ket{\psi(t)} = c_-(t)\ket{\phi_{-}(\vec{\theta}_t)} + c_+(t)\ket{\phi_{+}(\Bt_t)},
    \end{multline}
    The populations of the two states $|c_{\pm}(t)|^2$ vary more slowly as compared to the states $\ket{\phi_{\pm}(\vec{\theta}_t)}$. The states \(\ket{\phi_{\pm}(\vec{\theta}_t)}\) can be dressed by corrections controlled by \(|\vec{\omega}|/B_0\), such that the dressed state populations vary still more slowly.

    In the adiabatic limit the qubit mediates a quantised energy current.  The pumped power operator from drive 1 to drive 2 is $P_{1 \to 2} = \omega_1\partial_{\theta_1} H$. As the long time value of \([P_{1\to 2}]_t\) is \(\mathcal{O}(|\vec{\omega}|^2)\) (Eq.~\eqref{eq:pump_drive1_to2}), one must use the states $\ket{\tilde{\phi}_{\pm}(\vec{\theta}_t)}$ dressed to order $|\vec{\omega}|$. Evaluating the pumped power operator in the diagonal ensemble of the dressed basis gives~\cite{PhysRevLett.125.100601_PC_IM_AC}
    \begin{align}
        \expval{P_{1\to 2}} = \left( \omega_1 \partial_{\theta_1} |\vec{B}| + \omega_1 \omega_2 \tilde{\mathcal{B}} \right) \tilde{B}\cdot \expval{\vec{\sigma}} +\mathcal{O}\left(|\vec{\omega}|^3/B_0\right), 
\label{eq:pumped_power_expval_second_line}
    \end{align}
    where $\tilde{\mathcal{B}}(\vec{\theta}_t) = 2 \Im{\langle\partial_{\theta_1} \tilde{\phi}_{-}(\vec{\theta}_t)|\partial_{\theta_2} \tilde{\phi}_{-}(\vec{\theta}_t)\rangle}$ is the Berry curvature of the dressed ground state regarded as a function of $\vec{\theta}$ and $\tilde{B}\cdot \expval{\vec{\sigma}} = |\tilde{c}_-|^2 - |\tilde{c}_+|^2$ is the dressed state occupation difference. The dressed state occupation difference $\tilde{B}\cdot \expval{\vec{\sigma}}$ is a sum of two terms: the contribution from the instantaneous states $\hat{B}\cdot\expval{\vec{\sigma}}$, which we define as the \emph{spin lock fidelity}, and a rapidly oscillating term, which averages to $0$ at times $t \gg 1/|\vec{B}(t)|$.

    For an initial state prepared in an instantaneous eigenstate (so that \(\hat{B}\cdot\expval{\vec{\sigma}(0)} = \pm 1\)), the time averaged energy current follows from averaging Eq.~\eqref{eq:pumped_power_expval_second_line},
    \begin{equation}
    \begin{split}
        [\expval{P_{1\to 2}}]_t &\sim \left(\omega_1 \left[\partial_{\theta_1} |\vec{B}(\vec{\theta}_t)|\right]_t +  \omega_1 \omega_2 \left[\tilde{\mathcal{B}}(\vec{\theta}_t)\right]_t \right) \\
        &\sim C\frac{\omega_1 \omega_2}{2\pi}.
        \label{eq:pumped_power_der}
        \end{split}
    \end{equation}
Here, $[\cdot]_t$ denotes an average \([X]_t = \frac{1}{t}\int_0^t X(t') \d t'\), and $\sim$ denotes asymptotic equality in the limit of $t\to \infty$ such that $T_1 \ll t \ll \tau$. $C$ is the Chern number associated with the topology of the dressed eigenstates---equivalent to the Chern number of the corresponding instantaneous eigenstates~\eqref{eq:inst_eigenstates}. The first term in the upper line of Eq.~\eqref{eq:pumped_power_der} averages to 0 at times $t \gg T_1$. The lower line in Eq.~\eqref{eq:pumped_power_der} follows directly from integrating the Berry curvature on the $\vec{\theta}$ torus.

Therefore, in the adiabatic limit, the average energy current is quantised in the units of the Chern number of the instantaneous eigenstates
\begin{equation}\label{eq:pump_power}
    P_q = C\frac{\omega_1 \omega_2}{2\pi}.
\end{equation}

\subsection{Non-adiabatic effects}

For any finite drive frequency, and in the absence of fine-tuning~\cite{PhysRevB.99.064306_PC_IM_AC,Sels2017}, at late times non-adiabatic effects become important. This causes the time averaged energy current to vanish;
\begin{equation}\label{eq:net_power}
    \lim_{t \to \infty}\left[\expval{P_{1\to2}}\right]_t = 0.
\end{equation}
Equation~\eqref{eq:net_power} follows from results regarding the behaviour of generic two tone driven qubit models~\cite{PhysRevB.99.064306_PC_IM_AC,PhysRevB.105.144204_MBL_QP}---they only support topologically trivial steady states.

The energy current deviates from its quantized value due to variations in spin lock fidelity, which can be modelled by transitions between instantaneous eigenstates of the qubit. The probability of transition between the states (per period of one of the drives) is given by the exponentially small Landau-Zener (LZ) transition probability $p_{\mathrm{LZ}} = e^{-2 \pi \delta}$. It thus takes exponentially many periods to produce an \(O(1)\) probability of excitation, and a reversal of the energy current. 
In later sections we use a numerical \emph{transfer matrix} technique to show that $\expval{P_{1\to2}}$ is a function oscillating on timescale $\mathcal{O}(\tau)$.

\section{Adiabatic impulse model}\label{sec:num_model_TM}

In this section we develop an \emph{adiabatic impulse model} (AIM)~\cite{zener1932non,shevchenko2010landau,ivakhnenko2023nonadiabatic,kuno2019non_LZ_intf,lim2014mass_LZ_intf,lim2015geometric_LZ_intf,PhysRevB.97.035428_LZ_intf,tomka2018accuracy_LZ_intf,PhysRevA.73.063405_LZ_intf} for calculating the energy current efficiently, allowing us to access the exponentially long times necessary to observe reversal of the current. In Sec.~\ref{sec:cur_proxy} we relate the average energy current to the spin lock fidelity, which can be calculated with the transfer matrix method of Sec.~\ref{sec:trans_mat}. 

\subsection{Energy current proxy}\label{sec:cur_proxy}

The average spin lock fidelity is a good proxy for the average energy current mediated by the qubit. This follows from averaging the pumped power, given by Eq.~\eqref{eq:pumped_power_expval_second_line}, on the timescales $s \gg T_1$
    \begin{equation}\label{eq:spin_lock_pump_power}
    \mathrm{SA} \left[\expval{P_{1\to2}} \right]_s \left(t_0\right) = P_q \, \mathrm{SA} \left[ \hat{B}\cdot \expval{\vec{\sigma}} \right]_s \left(t_0\right) + \mathcal{O}(s^{-1}), 
    \end{equation}
    where 
    \begin{equation}\label{eq:sliding_avg_def}
        \mathrm{SA} \left[ X \right]_s\!(t_0) = \frac{1}{s} \int_{t_0}^{t_0 + s} dt \, X(t)
    \end{equation}
    is a sliding average.

    The key assumption necessary for the validity of Eq.~\eqref{eq:spin_lock_pump_power} is the separation of timescales on which the different components of the two terms in Eq.~\eqref{eq:pumped_power_expval_second_line} vary. The total derivative of the qubit's energy averages to $0$ on timescales $s =\mathcal{O}(T_1)$. The Berry curvature term averages to $P_q$ on timescales set by the convergence of its line integral to an area integral. This occurs when $s=\mathcal{O}(c T_1)$, with a constant $c$ controlled by the smoothness of the Berry curvature $\tilde{\mathcal{B}}(\vec{\theta})$ on the $\vec{\theta}$ torus. The spin lock fidelity of the dressed states, \(\tilde{B}\cdot \expval{\vec{\sigma}}\) has a slowly varying component given (to first order in \(|\vec{\omega}|\)) by \(\vec{B}\cdot \expval{\vec{\sigma}}\), and an additional small, rapidly oscillating component (due to the initial state being a bare eigenstate, and not a dressed state). The latter averages to zero on timescales \(s = \mathcal{O}(1/|\vec{B}|)\), while the former only varies on times exponential in the adiabatic parameter $\tau =\mathcal{O}\left(e^{2 \pi \delta} T_1\right)$. Therefore, the required separation of timescales justifying Eq.~\eqref{eq:spin_lock_pump_power} is satisfied in the adiabatic limit, where we have \(1/|\vec{B}|\ll T_1 \ll \tau\).

\begin{figure}
        \centering
        \includegraphics[width=\linewidth]{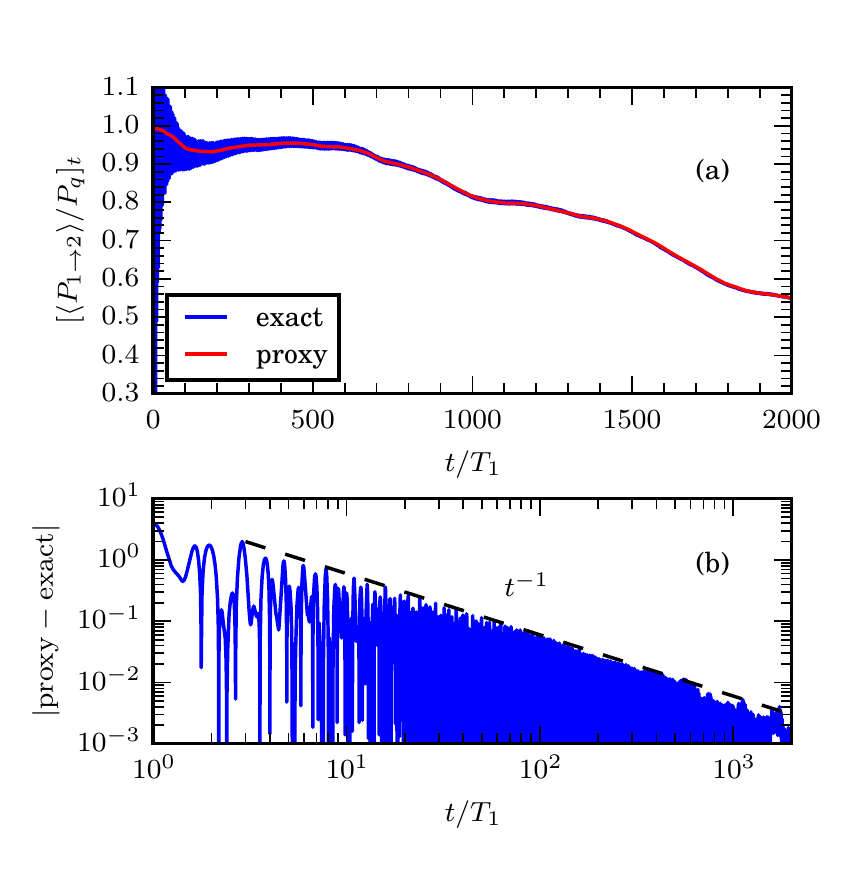}
        \caption{(a)~Time averaged energy current calculated through numerical integration $[\expval{P_{1\to2}}]_t/P_q$ (exact) and via the spin lock fidelity $\big[\hat{B}\cdot \expval{\vec{\sigma}}\big]_t$ (proxy) for a single initial phase $\vec{\theta}_0 = (4.0321, 2.0645)$. (b)~Error in using proxy for average energy current measurement behaves as in Eq.~\eqref{eq:spin_lock_pump_power}. \emph{Parameters:} $B_0=2$, $A=1$, $T_1 = 35$ in the BHZ model Eq.~\eqref{eq:normal_bhz_model}.}
        \label{fig:fig2}
    \end{figure}

    Figure~\ref{fig:fig2}(a) shows the time averaged energy current calculated exactly and approximately via the spin lock fidelity. Time averaging is equivalent to $t_0=0$ in Eq.~\eqref{eq:sliding_avg_def}. The difference between the exact and approximate curves (Fig.~\ref{fig:fig2}(b)) confirms the error terms in Eq.~\eqref{eq:spin_lock_pump_power} indeed decrease as \(\mathcal{O}(s^{-1})\). Similar evidence for the skew BHZ model with $A \gg 1$ is shown in Appendix~\ref{sec:energy_current_proxy_skew}.

\subsection{Transfer matrix evolution}\label{sec:trans_mat}
Non-adiabatic processes are the most significant when the energy gap between the instantaneous states is small compared to its typical value. By approximating time evolution as perfectly adiabatic away from such avoided crossings, and treating non-adiabatic transitions near the crossings as instantaneous, we arrive at a transfer matrix representation of the evolution operator. This approximation is known as the adiabatic impulse model (AIM)~\cite{zener1932non,shevchenko2010landau,ivakhnenko2023nonadiabatic,kuno2019non_LZ_intf,lim2014mass_LZ_intf,lim2015geometric_LZ_intf,PhysRevB.97.035428_LZ_intf,tomka2018accuracy_LZ_intf,PhysRevA.73.063405_LZ_intf}.

The AIM approximation to the evolution of the qubit from time \(0\) to \(t\) is given by 
\begin{equation}\label{eq:AIM}
\begin{split}
    U_{\mathrm{AIM}}(t, 0) = G(t, t_N)M(t_N)G(t_N, t_{N-1}) \dotsc \\
    \dotsc G(t_2, t_1)M(t_1)G(t_1, 0),
\end{split}
\end{equation}
where there are $N$ avoided level crossings given by the instantaneous energy minima $|\vec{B}(t_i)| = 0$, $\partial_t^2 |\vec{B}(t_i)| > 0$. 

In the basis of instantaneous eigenstates of the Hamiltonian, the adiabatic part of the evolution is expressed as
\begin{equation}
    G(t_{i+1}, t_i) = \e^{-i \sigma_z \xi(t_{i+1}, t_i) }.
\end{equation}
This captures the phase accrued during the adiabatic evolution between consecutive avoided level crossings, $t=t_i$ and $t=t_{i+1}$. This phase consists of a dynamical and geometric part, 
\begin{equation}\label{eq:adiabatic_phase}
    \xi(t_{i+1}, t_i) = \int_{t_i}^{t_{i+1}} dt \, \left[\frac{|\vec{B}(t)|}{2} - \matrixel{\phi_+(t)}{i\partial_t}{\phi_+(t)}\right],
\end{equation}
corresponding to the first and second terms in the integrand respectively. The geometric phase depends on the gauge choice of $\ket{\phi_\pm(t)}$. For explicit calculations, we pick the north pole gauge: 
\begin{align}\label{eq:gauge_choice}
    \ket{\phi_{+}(t)} &= \begin{pmatrix}
    -\sin(\eta(t)/2) \\
    e^{i\chi(t)} \cos(\eta(t)/2)
    \end{pmatrix} \\
    \ket{\phi_{-}(t)} &= \begin{pmatrix}
    e^{-i\chi(t)} \cos(\eta(t)/2) \\
    \sin(\eta(t)/2)
    \end{pmatrix},
\end{align}
where the coordinates $(\eta, \chi)$ are the spherical coordinates for \(\hat{B}(t)\).

The transfer matrix $M\left(t_i\right)$ captures the transition amplitudes between the instantaneous eigenstates at the avoided level crossing at $t=t_i$. For the skew BHZ model ($A \gg 1$ in Eq.~\eqref{eq:normal_bhz_model}), this matrix is
\begin{equation}\label{eq:tm_general}
    M(t_i) = \begin{pmatrix}
    e^{-i \phi_s(t_i)}\sqrt{1-p_{\mathrm{LZ}}(t_i)} & -e^{i \nu(t_i)} \sqrt{p_{\mathrm{LZ}}(t_i)} \\
    e^{-i \nu(t_i)}\sqrt{p_{\mathrm{LZ}}(t_i)} & e^{i \phi_s(t_i)}\sqrt{1-p_{\mathrm{LZ}}(t_i)}
    \end{pmatrix},
\end{equation}
where 
\begin{equation}\label{eq:trans_prob}
        p_{\mathrm{LZ}}(t_i) = e^{-2\pi \delta(t_i)}
\end{equation}
$\phi_s(t_i) = \pi/4 + \arg{\Gamma(1-i\delta(t_i))} + \delta(\log{\delta(t_i)} - 1)$, and $\Gamma(x)$ is the Gamma function. The adiabatic parameter at each avoided level crossing is 
\begin{equation}\label{eq:ad_param_ti}
    \delta(t_i) = \frac{|\vec{B}(t_i)|^2}{4 |\partial_t \vec{B}(t_i)|}.
\end{equation}
The transition probability~\eqref{eq:trans_prob} is largest when the adiabatic parameter $\delta(t_i)$ is smallest. The minima of Eq.~\eqref{eq:ad_param_ti} occur at $\cos(\theta_{1t_i})=1$ in Eq.~\eqref{eq:normal_bhz_model}, and are given by
\begin{equation}\label{eq:skew_lz_probability}
    \delta = \frac{B_0}{4A\omega_1}\left( 1 + \mathcal{O}(A^{-1})\right).
\end{equation}
The phase $\nu(t_i)$ is fixed by the gauge choice for the adiabatic states. Appendix~\ref{app:off-diag-phase} contains a detailed derivation of Eq.~\eqref{eq:tm_general}, and the function $\nu(t_i)$.

\begin{figure}
        \centering
        \includegraphics[width=\linewidth]{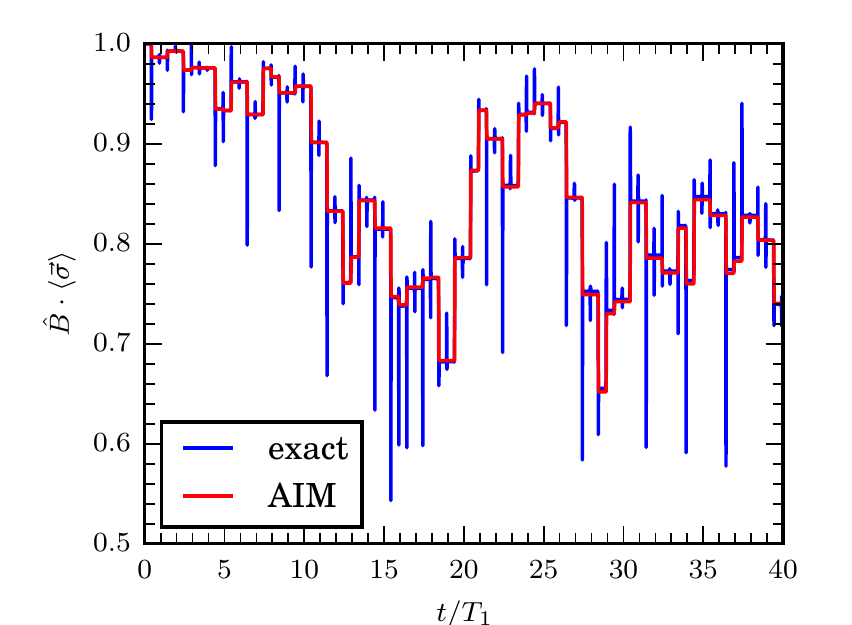}
        \caption{Instantaneous spin lock fidelity for a single initial state and phase $\Vec{\theta}_0$  calculated with the transfer matrix (AIM) and direct integration (exact). Note the ringing effects occurring at each crossing point which are not captured with AIM. These become unimportant upon averaging over time or initial phases. \emph{Parameters:} $B_0=2$, $A=30$, $T_1 = 300$, and $\Vec{\theta}_0=(3.6223, 0.9714)$ in the skew BHZ model Eq.~\eqref{eq:normal_bhz_model}.}
        \label{fig:Fig3-TM/intg_comparison}
    \end{figure}

Fig.~\ref{fig:Fig3-TM/intg_comparison} shows that AIM accurately predicts the average adiabatic state population for the skew BHZ model~\eqref{eq:normal_bhz_model}. When $A \gg 1$, the adiabatic parameter of the drive $\delta(t)$ has well separated local minima---making it well suited to treatment with AIM. The instantaneous adiabatic populations show ringing around the crossing points which the transfer matrix does not capture. However, such oscillations have a negligible contribution to average quantities, including the average spin lock fidelity and average energy transferred between the drives.

\section{Giant energy oscillations}\label{sec:results}

The total energy transferred from drive 1 to drive 2
\begin{equation}\label{eq:energy_transfer}
    \Delta E_2(t) = \int_0^t \, dt' \expval{P_{1\to2}},
\end{equation}
is bounded, but \emph{giant} in comparison to the qubit bandwidth $AB_0$. This result follows from the slow oscillations of $\mathrm{SA}\left[\expval{P_{1\to2}}\right]$, which integrate to large amplitude excursions in~\eqref{eq:energy_transfer}. 

Numerically verifying this claim requires accurately calculating the oscillatory function $\expval{P_{1\to2}}$ over very long timescales. Sec.~\ref{sec:energy_slosh} discusses the results of doing this for the skew BHZ model~\eqref{eq:normal_bhz_model} using AIM and the spin lock fidelity. We find that the typical timescale of energy oscillations $\tau$ is a non-monotonic function of the adiabatic parameter $\delta$ (Sec.~\ref{sec:period_od_slosh}). We interpret this as a result of interference effects in the transition amplitudes. Introduction of decoherence in qubit dynamics reduces the scale of non-monotonic dependence of $\tau$ with $\delta$ (Sec.~\ref{sec:decoherence}); providing evidence for our claim. 

\subsection{Energy current oscillations}\label{sec:energy_slosh}

\begin{figure}
    \centering
    \includegraphics[width=\linewidth]{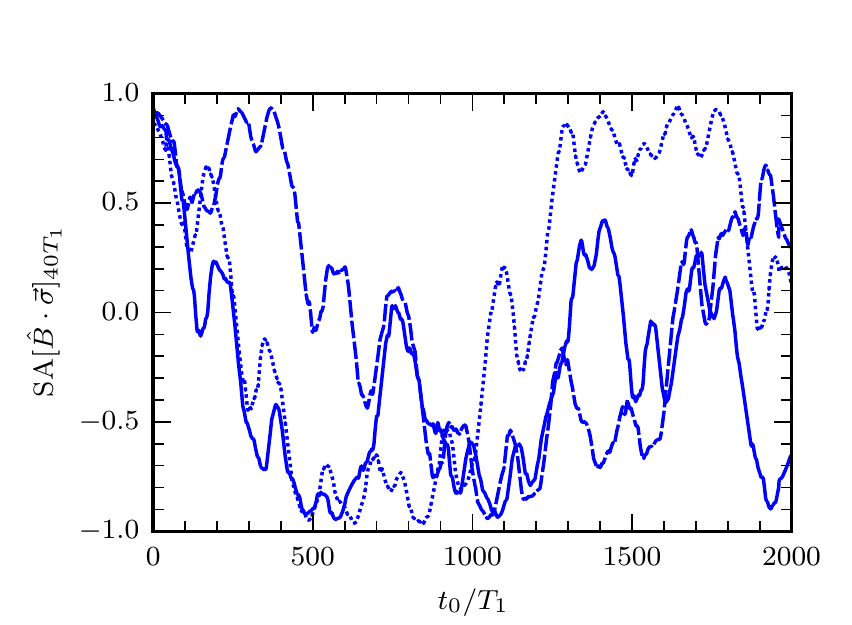}
    \caption{Oscillations in the average energy current via the sliding average ($s=40T_1$) of spin lock fidelity using AIM in three initial phases: full $\vec{\theta}_0=(1.4454, 5.3288)$, dashed $\vec{\theta}_0=(0.2324, 2.6727)$ and dotted $\vec{\theta}_0=(1.6707, 1.6094)$. \emph{Parameters:} $B_0=2$, $A=30$, $T_1 = 320$ in the skew BHZ model Eq.~\eqref{eq:normal_bhz_model}.}
    \label{fig:energy_sloshing}
\end{figure}

The qubit mediates a quasiperiodically oscillating energy current for each initial phase $\vec{\theta}_0$ (Fig.~\ref{fig:energy_sloshing}). The energy current reverses direction approximately every $\tau$, capturing the timescale of energy current oscillations.

Energy current oscillations result in quasiperiodic oscillations of the energy transferred into drive two, $\Delta E_2$, with an amplitude set by the typical timescale of the energy current oscillations (Fig.~\ref{fig:2_tone_driven_qubit_sketch}) 
\begin{equation}
    \Delta E_{2,\max} = \mathcal{O}(P_q\tau).
\end{equation}
The timescale of energy current oscillations $\tau$ stems from non-adiabatic effects in the qubit dynamics. The probability of transition between instantaneous eigenstates per avoided crossing is given by $p_{\mathrm{LZ}} = e^{-2\pi \delta}$, with $\delta$ given by \eqref{eq:skew_lz_probability}. As there are $\mathcal{O}(1)$ avoided crossings per period $T_1$, the transition rate is $\mathcal{O}(p_{\mathrm{LZ}}/T_1)$. Ignoring the coherence effects between subsequent avoided crossings, this calculation predicts $\tau = \mathcal{O}(T_1/p_{\mathrm{LZ}})$. Thus, the scale of energy current oscillations is exponentially large in the adiabatic parameter~\eqref{eq:skew_lz_probability}, 
\begin{equation}
    \Delta E_{2,\max} = \mathcal{O}(e^{2\pi \delta}\omega_2).
\end{equation}
In the adiabatic limit of the drive this energy mediated by the qubit is much larger than the bandwidth of the qubit $A B_0$---the oscillations are \emph{giant}. 

Quasiperiodic energy current oscillations follow from previous work. Reference~\cite{PhysRevB.99.064306_PC_IM_AC} shows that a generic \(d\)-level system driven by two tones at finite frequency exhibits a pure-point spectrum. This means that generic observables exhibit coherent quasiperiodic oscillations with a finite set of fundamental frequencies (three for a qubit).

The energy current oscillation patterns in Fig.~\ref{fig:energy_sloshing} are very sensitive to initial phases of the drive at times $t > \tau$. More precisely, the different initial phases have correlated energy currents on timescales $t < \tau$,
\begin{equation}\label{eq:spread_in_traj_small_t}
    \langle P_{1\to2}(t, \Vec{\theta_0}) \rangle - \langle P_{1\to2}(t, \Vec{\theta_0}') \rangle = \mathcal{O}(t/\tau)
\end{equation}
for $(\Vec{\theta_0} - \Vec{\theta_0}') \cdot \vec{\nabla}_{\vec{\theta}_0} \mathcal{B}(\vec{\theta}_0) \ll 1$, where $\mathcal{B}(\vec{\theta})$ is the Berry curvature of the instantaneous ground state defined on the $\vec{\theta}$ torus. Beyond this time, $\langle P_{1\to2}(t, \Vec{\theta_0}) \rangle$ remains a continuous function of $\Vec{\theta_0}$ but is only smooth on initial phase separation scales inversely proportional to $\tau$, $|\Vec{\theta}_0-\Vec{\theta}_0'| < \mathcal{O}((\tau/T_1)^{-1})$ (Appendix~\ref{sec:sensitivity_to_init_phase}).

\subsection{Typical timescale of energy oscillations}\label{sec:period_od_slosh}
\begin{figure}
    \centering
    \includegraphics[width=\linewidth]{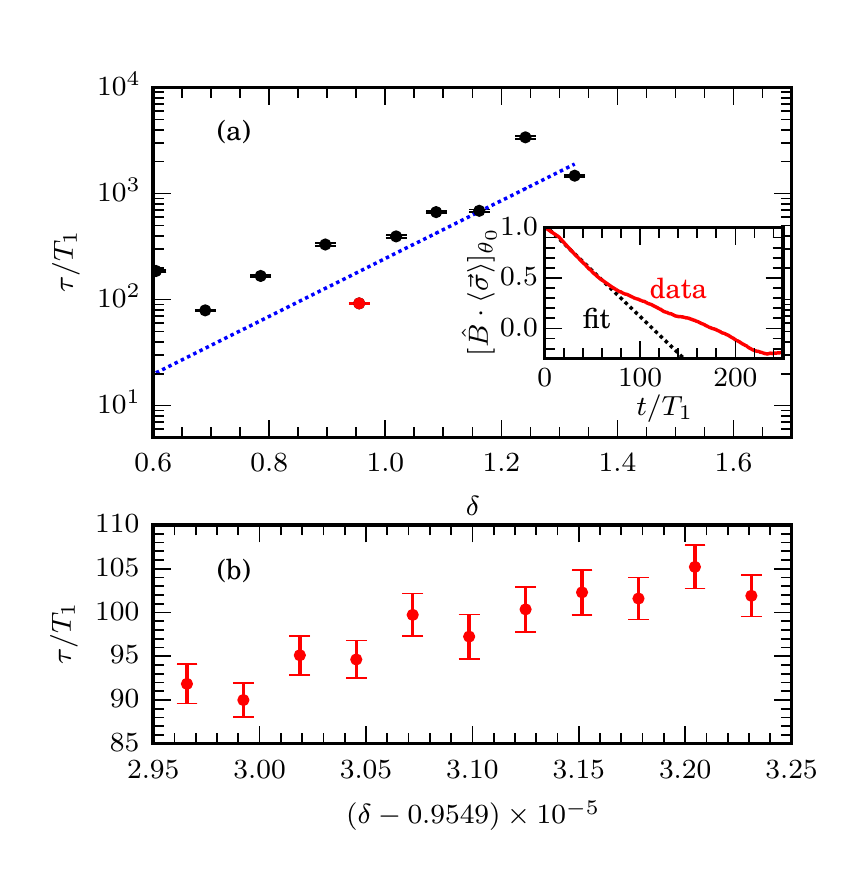}
\caption{(a)~The typical timescale \(\tau\) of energy current oscillations (black points) scale as \(a e^{2\pi \delta} T_1\) (blue line), with large deviations. The adiabaticity parameter \(\delta\) is calculated from Eq~\eqref{eq:skew_lz_probability} with model \emph{parameters} $B_0 = 2$, $A=30$ and varying $T_1$. Timescales \(\tau\) are measured by fitting early time average (denoted by $[\cdot]_{\vec{\theta}_0}$, over 1000 initial phases) spin lock fidelity to \(b-t/\tau\) between $ 0.4 \lesssim \hat{B}\cdot \langle \vec{\sigma} \rangle < 1$ (inset). Errors are estimated through bootstrap re-sampling. (b)~The scale in the variation of \(\delta\) such that $\tau$ is smooth is very small, $\Delta \delta \approx 10^{-7}$. $\tau$ is measured as in (a), only with 500 initial phase realizations.}
    \label{fig:pumping_lifetime_std_ad_param}
\end{figure}

    Numerically measuring $\tau$ confirms its exponential scaling with the adiabatic parameter $\delta$ (Sec.~\ref{sec:energy_slosh}). However, the detailed dependence of $\tau(\delta)$ is non-monotonic in the adiabatic parameter $\delta$ (Fig.~\ref{fig:pumping_lifetime_std_ad_param}(a)). 

    The timescale $\tau$ is estimated through a fit to the early time drop-off in the energy current. The initial phase averaged energy current has the form 
    \begin{equation}\label{eq:linear_deviation}
        [\expval{P_{1\to2}}]_{\Vec{\theta}_0} = P_q\left(1 - t/\tau +\mathcal{O}\left(\left(t/\tau\right)^2\right)\right),
    \end{equation}
    so a linear fit to the initial phase averaged spin lock fidelity (Fig.~\ref{fig:pumping_lifetime_std_ad_param}(a) inset) produces an estimate of $\tau$ (Fig.~\ref{fig:pumping_lifetime_std_ad_param}(a)). Averaging over initial phases captures the mean response of the energy current for every adiabatic parameter $\delta$. This is valid due to insensitivity of the energy current on the initial phase $\vec{\theta}_0$ at times $t<\tau$ (Fig.~\ref{fig:energy_sloshing}). 

    In Fig.~\ref{fig:pumping_lifetime_std_ad_param} (b), we observe variation between \(\tau(\delta)\) and \(\tau(\delta+\Delta\delta)\) at extremely small scales in the difference \(\Delta\delta\). We can estimate the \(\Delta\delta\) required to have \(\Delta \tau \ll T_1\) by assuming that this variation is due to the dynamical phase accrued by the adiabatic states, and thus is an effect of coherence in the dynamics. 
    
    Suppose we perturb the Hamiltonian parameters on the scale \(|\vec{B} - \vec{B}'| = \Delta B\), such that \(\tau\) changes to \(\tau'\). The difference in the dynamical phase accrued within the time \(\tau\) is \(\mathcal{O}(\Delta B \tau)\). When \(\Delta B \tau \ll 1\), dynamics is unaffected within time \(\tau\). As the evolution of any observable is quasiperiodic with an oscillation scale which is \(\mathcal{O}(\tau)\), the variation in an observable between the perturbed an unperturbed Hamiltonian being bounded within time \(\mathcal{O}(\tau)\) implies that the variation is bounded for all time. In particular, \(\vec{B'}\cdot\expval{\vec{\sigma}}'\) must remain close to \(\vec{B}\cdot\expval{\vec{\sigma}}\). The former defines the new \(\tau'\), so when
    \begin{equation}
        \Delta B \tau \ll 1 
        \quad\implies\quad
        \Delta \delta \ll e^{-2\pi\delta},
        \label{eq:ad_param_scale_of_smoothness}
    \end{equation}
    (where we used \(\Delta B \approx \Delta\delta/T_1\) and \(\tau = \mathcal{O}(T_1 e^{2\pi \delta})\)) it follows that \(|\tau'-\tau| \ll T_1\).

    Note that the change in the transition amplitude \(p_{\mathrm{LZ}}\) is exponentially smaller in \(\delta\) than \(\Delta B\). Thus, assuming dynamics is governed by the accrued phase, rather than the non-adiabatic crossings, estimates a much smaller value of \(\Delta \delta\) such that \(\tau\) is smooth. The smallness of this estimate conforms with our numerical observations (Fig.~\ref{fig:pumping_lifetime_std_ad_param}(b)). In fact, we observe variation in \(\tau\) at scales even smaller than \(e^{-2\pi\delta}\). Equation~\eqref{eq:ad_param_scale_of_smoothness} should be interpreted as a scaling estimate (because we used the scaling expression \(\tau = \mathcal{O}(T_1 e^{2\pi \delta})\)), so it is possible that the coefficient in this scaling happens to be very small.

    The derivation of the exponential scaling ignored coherence effects between subsequent avoided crossings. These coherence effects have a non-trivial role in determining the detailed dependence of $\tau$, and result in 
    \begin{equation}
        \tau \approx C(\delta, \phi_{\mathrm{ad}}) e^{2\pi \delta}T_1.
    \end{equation}
     Here, $C(\delta, \phi_{\mathrm{ad}})$ is set by the adiabatic parameter and phase $\phi_{\mathrm{ad}}$ accrued during adiabatic evolution of the qubit between avoided crossings. Interference effects can cause $C(\delta, \phi_{\mathrm{ad}})$ to have a non-monotonic dependence on $\delta$, consistent with our numerical observations. 

\subsection{Decoherence}\label{sec:decoherence}
To further test the hypothesis that the non-monotonicity of \(\tau\) is due to interference effects, we simulate the addition of dephasing to the qubit dynamics. This results in the decoherence in the instantaneous basis and, if our hypothesis is correct, results in \(\tau\) becoming a monotonic function of \(\delta\).

\begin{figure}
    \centering
    \includegraphics[width=\linewidth]{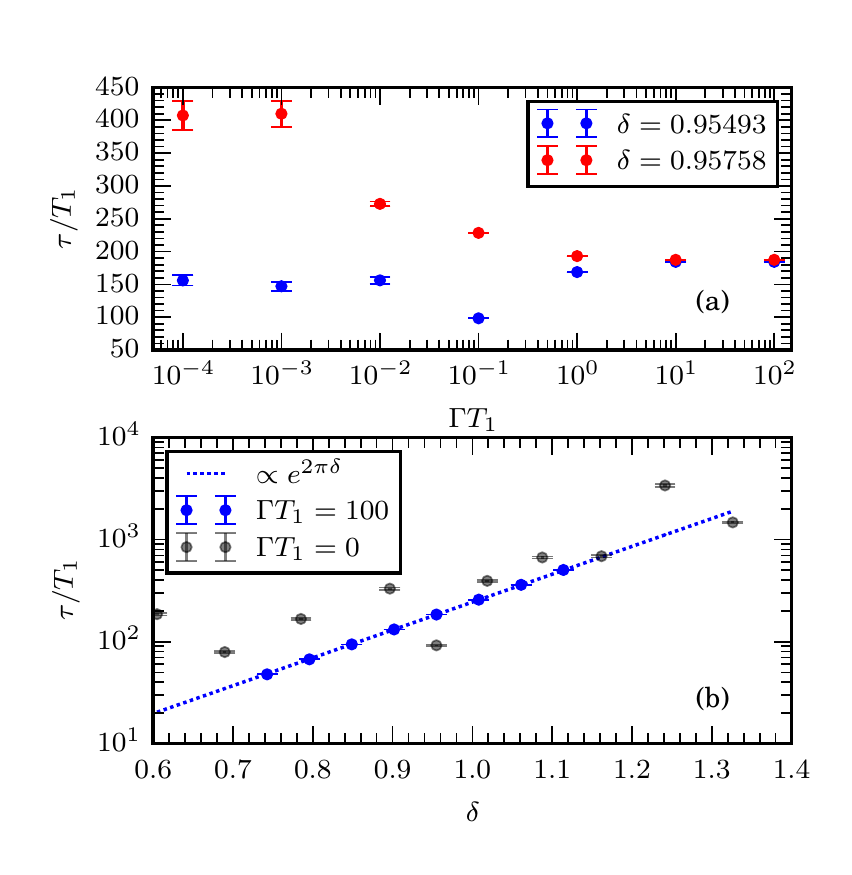}
    \caption{(a)~The sensitive dependence of \(\tau\) on \(\delta\) is eliminated by introducing dephasing of sufficient strength \(\Gamma\). The timescale \(\tau\) is measured is in Fig.~\ref{fig:pumping_lifetime_std_ad_param} for small \(\Gamma\) (which shows underdamped oscillations of \(\vec{B}\cdot\expval{\vec{\sigma}}\)), and as the timescale for the exponential decay of the spin lock fidelity for large \(\Gamma\) (where \(\vec{B}\cdot\expval{\vec{\sigma}}\) is overdamped). \emph{Parameters in model~\eqref{eq:normal_bhz_model}}: $B_0 = 2$, $A=30$, $T_1=360$ (blue curve) and $T_1=361$ (red curve) with averages over 200 initial phase realizations. (b)~Sufficient dephasing causes \(\tau\) to become a monotonic function of \(\delta\) (blue). The light gray points are a copy of data in Fig.~\ref{fig:pumping_lifetime_std_ad_param}(a), and the blue line is proportional to \(e^{2\pi\delta}\) (same as the blue line in Fig.~\ref{fig:pumping_lifetime_std_ad_param}(a)).}
    \label{fig:lifetime_dephasing}
\end{figure}

Indeed, in Fig.~\ref{fig:lifetime_dephasing}(b) we observe that sufficiently strong decoherence causes $\tau$ to become monotonic. The noise model is implemented in the AIM evolution of the qubit (Sec.~\ref{sec:trans_mat}) via a quantum channel~\cite[Chapter 8]{nielsen00} (in the Krauss formulation)---representing noise in the dynamical phase of the qubit evolution. Specifically, we make a replacement in the adiabatically accrued phase~\eqref{eq:adiabatic_phase} 
\begin{equation}\label{eq:noise}
\xi(t_{i+1}, t_i) \to \xi(t_{i+1}, t_i)+\int_{t_{i}}^{t_{i+1}} dt \, \eta(t), 
\end{equation}
where are $\eta(t)$ are i.i.d. normal random variables with the two point correlator $[\eta(t_{i+1})\eta(t_i)]_\eta = \Gamma \delta(t_{i+1}-t_i)$. Here $[x]_{\eta}$ denotes averaging over the Gaussian ensemble and $\delta(x)$ is the delta function. Averaging over the random noise processes in Eq.~\eqref{eq:noise} yields the effective AIM evolution for the density matrix  given by 

\begin{equation}\label{eq:dephasing_evo}
    \begin{split}
        \rho (t_{i+1}) = \sum_{j=+,-} K_j(t_{i+1}, t_{i}) G(t_{i+1}, t_{i}) M(t_i) \rho(t_i) \\ 
        M^{\dagger}(t_i) G^{\dagger}(t_{i+1}, t_{i}) K^{\dagger}_j(t_{i+1}, t_{i}),
    \end{split}
\end{equation}
where $\rho (t_{i+1})$ is the density matrix just before the avoided level crossing at time $t=t_{i+1}$, $G$ and $M$ are the adiabatic evolution and transfer matrices (Sec.~\ref{sec:trans_mat}). The Krauss operators $K_{\pm}$ are
\begin{equation}
    K_{\pm} = \frac{1}{\sqrt{2\left( 1+ \gamma^2 \right)}} \left( \mathbf{1} \pm \gamma \sigma_z \right),
\end{equation}
where $\gamma = \sqrt{\tanh{\left(\Gamma\left(t_{i+1}-t_{i}\right)/4\right)}}$.

At any non-zero decoherence rate $\Gamma$, the qubit mediates damped energy oscillations between the drives. The timescale $\tau$ remains the oscillatory timescale of the energy current in the underdamped regime $\Gamma < \tau^{-1}(\Gamma=0)$. However, in the overdamped regime $\tau$ should be interpreted as the exponential decay timescale of the energy current. Figure~\ref{fig:lifetime_dephasing}(a) shows that the values of $\tau$ for two nearby values of the adiabatic parameter $\delta$ converge to essentially the same limit at large decoherence rate---consistent with an incoherent model of population transfer at subsequent avoided level crossings.

\section{Discussion}
Our results show that a quasiperiodically driven qubit can mediate a slowly oscillating energy current between the drives, resulting in giant amplitude oscillations in the transferred energy. The typical timescale of energy current oscillations $\tau$ is exponentially large in the adiabatic parameter $\delta$, with non-monotonicities on small scales. 

Nitrogen-vacancy (NV) centers provide a room temperature platform for experimentally realizing giant energy oscillations. Indeed, Boyers et al.~\cite{PhysRevLett.125.160505_EB_PC_AC_AS} have already observed the topological regime ($C\neq 0$) of qubit dynamics. In this experiment, perfect adiabaticity was achieved via counterdiabatic driving---a fine tuned drive protocol which suppresses transitions between the instantaneous states~\cite{Sels2017}. Energy current oscillations will occur when the driving protocol is perturbed away from perfect counterdiabaticity; these can be indirectly measured through the spin-lock fidelity. The challenge is decoherence. The observed decoherence time $T_2 \approx 10\, \mu \mathrm{s}$ of the NV center~\cite{PhysRevLett.125.160505_EB_PC_AC_AS} requires a drive of frequency $\Omega \approx 10 \mathrm{MHz}$ and a perturbation of the same scale to see an energy current reversal.

Our results are relevant to other adiabatic topological pumps, including the disordered Thouless charge pump~\cite{nakajima2021competition_thouless_dis,citro2023thouless,cerjan2020thouless, PhysRevA.103.043310_thouless, PhysRevLett.120.106601_thouless}. Indeed, the synthetic lattices of the two tone driven qubit and the Thouless pump are closely related. Our results imply that at any non-zero frequency and disorder strength, charge pumping only persists for a finite time~\cite{dis_thouless_pump_future_work}. The total charge pumped would similarly be finite, but giant.

We showed that the non-monotonic behaviour of $\tau(\delta)$ arises from interference effects in qubit dynamics. As $\delta \to \infty$, $\tau(\delta)$ develops structure at exponentially small scales in $\delta$. The possibility of an underlying fractal structure is an intriguing avenue for future work~\cite{PhysRevB.14.2239_hoft, PhysRevE.96.032130_fractality}.

\section*{Acknowledgements}
The authors thank C. Baldwin, S. R. Koshkaki, M. Kolodrubetz, C. Laumann, E. McCulloch and A. Polkovnikov for helpful discussions. This work was supported by: NSF Grant No.DMR-1752759, and AFOSR Grant No. FA9550-20-1-0235 (D.V., D.L. and A.C.); and the NSF STC “Center for Integrated Quantum Materials” under Cooperative Agreement No. DMR-1231319 (P.C.). Numerical work was performed on the BU Shared Computing Cluster, using \textsc{Quspin}~\cite{10.21468/SciPostPhys.2.1.003_quspin, 10.21468/SciPostPhys.7.2.020_quspin}.

\appendix

\section{Derivation of the general transfer matrix}\label{app:off-diag-phase}

A change of basis in the model~\eqref{eq:normal_bhz_model} relates the transfer matrix $M\left(t_i\right)$ in Eq.~\eqref{eq:AIM} to the transfer matrix of the analytically solved Landau-Zaener (LZ) ramp Hamiltonian~\cite{zener1932non,ivakhnenko2023nonadiabatic,shevchenko2010landau}.

The LZ ramp Hamiltonian is given by 
\begin{equation}\label{eq:lz_ramp}
    H_{\mathrm{LZ}} = -\frac{vt}{2} \sigma_z - \frac{\Delta}{2} \sigma_x.
\end{equation}

In the adiabatic basis in north pole gauge, Eq.~\eqref{eq:gauge_choice}, the transfer matrix is
\begin{equation}\label{eq:trans_mat_lz_prob}
    M_{\mathrm{LZ}}=\begin{pmatrix}
    e^{-i \phi_s}\sqrt{1-p_{\mathrm{LZ}}} & -\sqrt{p_{\mathrm{LZ}}} \\
    \sqrt{p_{\mathrm{LZ}}} & e^{i \phi_s}\sqrt{1-p_{\mathrm{LZ}}}
    \end{pmatrix},
\end{equation}
where the parameters have the same form as in Eq.~\eqref{eq:skew_lz_probability}, but with the adiabatic parameter $\delta = \Delta^2/4v$. Note that the only variable controlling this matrix is $\delta$, which quantifies the adiabaticity of the ramp---a large $\delta$ gives an exponentially small probability of transition. The exact solution relies upon $H_{\mathrm{LZ}}$ being linear in \(t\).

A change of basis is necessary to compute the transfer matrix for a qubit driven by an arbitrary external field $H(t) = \vec{B}(t)\cdot \vec{\sigma}/2$. We expand the Hamiltonian near an avoided level crossing point $t=t_c$, defined as a minimum point of $|\vec{B}(t)|$.
\begin{equation}\label{eq:linearized_ham}
\begin{split}
    H(t-t_c) = \frac{1}{2}\Big(\vec{B}(t_c) \cdot \vec{\sigma} + (t-t_c)\partial_t \vec{B}(t_c) \cdot \vec{\sigma} \\
    + \mathcal{O}\big((t-t_c)^2\big)\Big)
\end{split}
\end{equation}

There is a unitary transformation which rotates the Hamiltonian~\eqref{eq:linearized_ham} into the LZ ramp Hamiltonian~\eqref{eq:lz_ramp}, allowing us to find \(M(t_c)\) for generic \(H(t)\), provided the quadratic error term of Eq.~\eqref{eq:linearized_ham} is small. This transformation exists because the two terms in Eq.~\eqref{eq:linearized_ham} are trace orthogonal when \(|\vec{B}(t_c)|\) is a local minimum,
\begin{equation}\label{eq:trace_orth}
    \partial_t \Tr[H(t)^2] = 2\Tr\big[(\vec{B}(t_c)\cdot \vec{\sigma}) (\partial_t \vec{B}(t_c) \cdot \vec{\sigma})\big] = 0.
\end{equation}
Thus, each term can be rotated into a distinct Pauli matrix with a unique (up to a phase) unitary transformation,
\begin{equation}\label{eq:rotation}
    H_{\mathrm{LZ}}(t) = U_{\mathrm{rot}}^{\dagger}(t_c)H(t-t_c) U_{\mathrm{rot}}(t_c).
\end{equation}
In sum, the transfer matrix is
\begin{equation}\label{eq:arbit_drive_trans_mat}
    M(t_c) = U_{\mathrm{rot}}(t_c)M_{\mathrm{LZ}}U_{\mathrm{rot}}^{\dagger}(t_c), 
\end{equation}
with matrix elements~\eqref{eq:tm_general} evaluated in the basis of Eq.~\eqref{eq:gauge_choice}.

The rotation matrix $U_{\mathrm{rot}}$ is constructed by lifting a corresponding SO(3) rotation matrix $R(t_c)$ to SU(2)---making $U_{\mathrm{rot}}$ unique up to a sign. The rotation matrix $R(t_c)$ is fixed by the linearized Hamiltonian at each avoided level crossing point $t_c$
\begin{equation}
    R(t_c) = \begin{pmatrix}
    \hat{B}(t_c) \\
    \partial_t \hat{B}(t_c) \cross \hat{B}(t_c) \\
    \partial_t \hat{B}(t_c)
    \end{pmatrix},
\end{equation}
where this construction ensures $\det(R(t_c)) = 1$ for all $t_c$. Describing $R(t_c)$ as a rotation by angle $\Theta$ around the axis of rotation $\hat{n}$, we lift it to SU(2) via $U_{\mathrm{rot}} = \exp(-i\Theta \hat{n} \cdot \Vec{\sigma}/2)$.

\begin{figure}
    \centering
    \includegraphics[width=\linewidth]{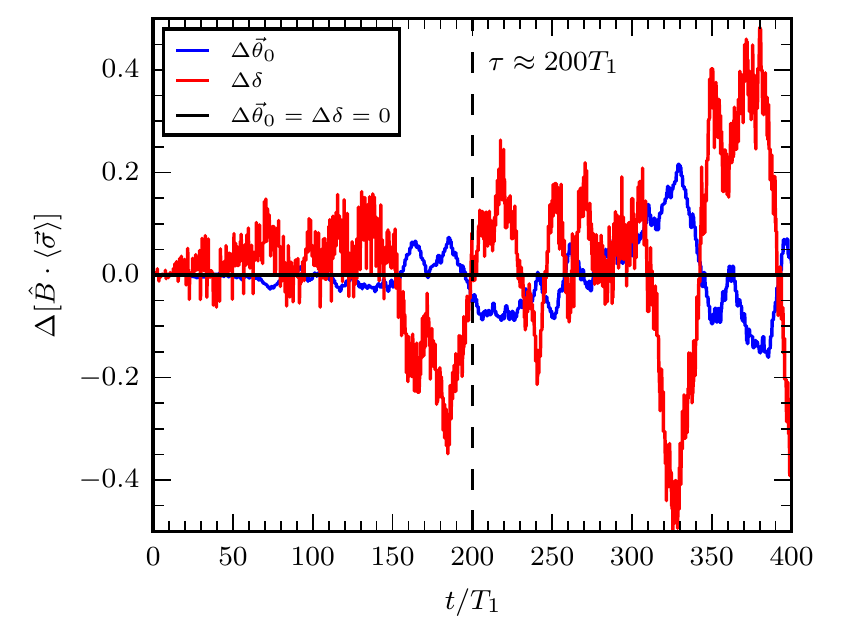}
    \caption{Deviation in the spin lock fidelity from the unperturbed set of initial conditions: $B_0=2$, $A=30$, $T_1 = 380$, $\vec{\theta}_0 = (2.4, 5.3)$ (black line). The perturbation in initial phase is given by $\vec{\theta}_0 = (2.401, 5.301)$ (blue line), with $|\Delta \vec{\theta_0}| = \mathcal{O}\left(10^{-3}\right)$. The perturbation in adiabatic parameter is given by a change in $T_1$ (red line), with $\Delta \delta = \mathcal{O}(10^{-7})$.}
    \label{fig:phase_delta_perturb}
\end{figure}

\begin{figure}
    \centering
    \includegraphics[width=\linewidth]{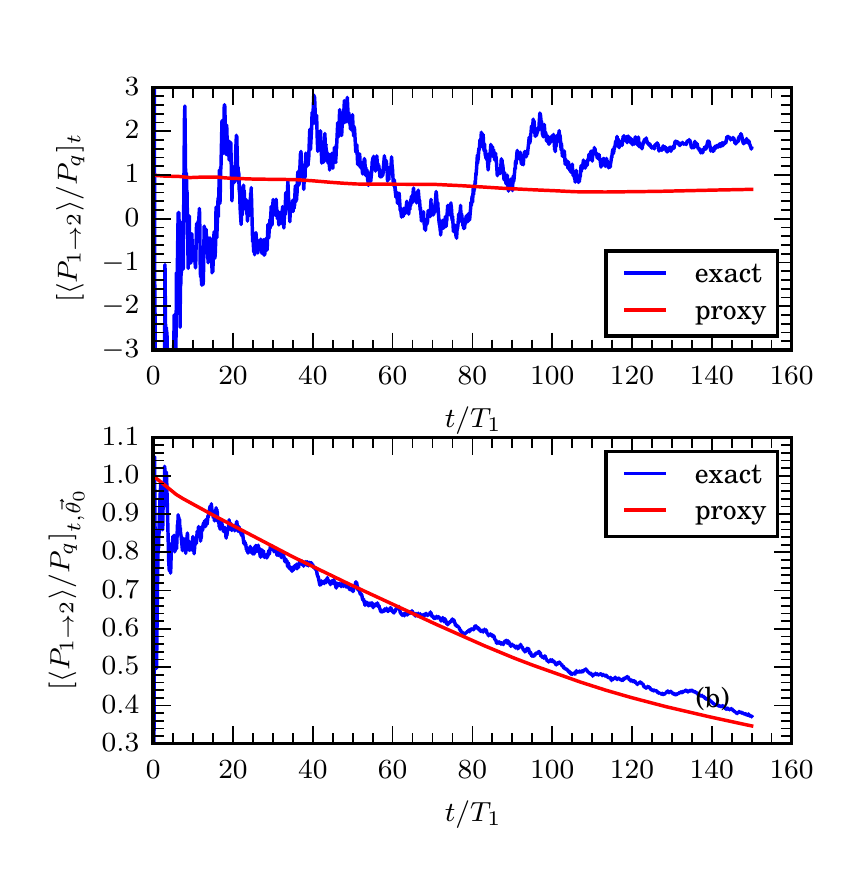}
    \caption{(a) Time averaged energy current calculated by numerical integration $[\expval{P_{1\to2}}]_t/P_q$ (exact) and via the spin lock fidelity, with AIM, $\big[\hat{B}\cdot \expval{\vec{\sigma}}\big]_t$ (proxy) for a single initial phase $\vec{\theta}_0 = (5.1545, 2.0452)$. (b) Initial phase averaged (denoted as in Fig.~\ref{fig:pumping_lifetime_std_ad_param}) curves (1000 realizations) for the \emph{parameters:} $B_0=2$, $A=30$, $T_1 = 300$ in the skew BHZ model Eq.~\eqref{eq:normal_bhz_model}. }
    \label{fig:proxy_skew_appendix}
\end{figure}

Matrix elements of the transfer matrix at each crossing point can now be related to the matrix elements of the transfer matrix for the LZ ramp problem (Eq.~\eqref{eq:trans_mat_lz_prob}) by inserting a rotation from the adiabatic states before the avoided crossing to those after, \(\sum_{kl} \ket{\phi^{\mathrm{LZ}}_k} \bra{\phi^{\mathrm{LZ}}_l}\), into
\begin{equation}
    M_{ij} (t_c) = \bra{\phi_i(t_c)} U_{\mathrm{rot}} M_{LZ} U^{\dagger}_{\mathrm{rot}} \ket{\phi_j(t_c)}.
\end{equation}
The diagonal elements remain unchanged as $U_{\mathrm{rot}}$ does not couple $k \neq i$ and $l \neq j$ elements. The off-diagonal elements acquire a phase difference $2\nu$ such that
\begin{equation}
    e^{i \nu(t_c)} = \left(\bra{\phi_{+}} U_{\mathrm{rot}} \ket{\phi_{+}^{\mathrm{LZ}}}\right)^2
\end{equation}
as denoted in Eq.~\eqref{eq:tm_general}.

Comparing the linearised Hamiltonian in Eq.~\eqref{eq:linearized_ham} with the LZ ramp problem Eq.~\eqref{eq:lz_ramp} we can identify the gap as $\Delta = |\vec{B}(t_c)|$ and the velocity $v=|\partial_t \vec{B}(t_c)|$---giving the adiabatic parameter at each avoided crossing as 
\begin{equation}
    \delta(t_c) = \frac{|\vec{B}(t_c)|^2}{4 |\partial_t \vec{B}(t_c)|}.
\end{equation}

\section{Sensitivity to perturbations}\label{sec:sensitivity_to_init_phase}

The energy current is exponentially sensitive to perturbations in the drive. This is a consequence of the long timescale $\tau$ on which the qubit reverses the energy current direction. The spin lock fidelity of the perturbed initial phase $\Delta \vec{\theta}_0 \le \mathcal{O}\left(e^{-2 \pi \delta}\right)$, or adiabatic parameter $\Delta \delta \ll \mathcal{O}\left(e^{-2 \pi \delta}\right)$ start deviating from the unperturbed spin lock fidelity around $t = \mathcal{O}(\tau)$. Indeed, Fig.~\ref{fig:phase_delta_perturb} shows the difference in the spin lock fidelity as a function of time between the unperturbed set of initial conditions (black line) and the perturbed ones. The difference in the spin lock fidelity starts growing significantly near $t = \mathcal{O}(\tau)$.

\section{Energy current proxy for the skew BHZ model}\label{sec:energy_current_proxy_skew}

The average energy current is well approximated by the average spin lock fidelity even in the skew BHZ model~\eqref{eq:normal_bhz_model} with $A \gg 1$. That is, Eq.~\eqref{eq:spin_lock_pump_power} still holds with a larger prefactor than in the BHZ model ($A=1$ in Eq.~\eqref{eq:normal_bhz_model}). This can be seen by comparing Fig.~\ref{fig:fig2} with Fig.~\ref{fig:proxy_skew_appendix}(a). In the skew BHZ mode,l the micromotion of the qubit is much larger---the maximum energy the qubit can absorb is proportional to $A B_0$. Averaging over this micromotion controls the scale of the error term in Eq.~\eqref{eq:spin_lock_pump_power}. Averaging over the initial phases of the drive in Eq.~\eqref{eq:normal_bhz_model} decreases the scale of the error term (Fig.~\ref{fig:proxy_skew_appendix}(b)). This is because the micromotion of the qubit is independently distributed for each initial phase of the drive.

\bibliography{ref.bib}

\end{document}